# Theory and experiment of isotropic electromagnetic beam bender made of dielectric materials

Qibo Deng[1], Jin Hu[2,a], Zheng Chang[1], Xiaoming Zhou[1] and Gengkai Hu[1,b]

[1] School of Aerospace Engineering, Beijing Institute of Technology, Beijing, 100081, P. R. China

[2] School of Information and Electronics, Beijing Institute of Technology, Beijing, 100081, P. R. China

[a]bithj@bit.edu.cn, [b]hugeng@bit.edu.cn

**Abstract.** In this paper, we utilize the deformation transformation optics (DTO) method to design electromagnetic beam bender, which can change the direction of electromagnetic wave propagation as desire. According to DTO, the transformed material parameters can be expressed by deformation tensor of the spatial transformation. For a beam bender, since the three principal stretches at each point induced by the spatial transformation are independent to each other, there are many possibilities to simplify the transformed material parameters of the bender by adjusting the stretches independently. With the DTO method, we show that the reported reduced parameters of the bender obtained by equivalent dispersion relation can be derived as a special case. An isotropic bender is also proposed according to this method, and it is fabricated by stacking dielectric materials in layered form. Experiments validate the function of the designed isotropic bender for a TE wave; it is also shown that the isotropic bender has a broadband with low loss, compared with the metamaterial bender. The isotropic bender has much easier design and fabrication procedures than the metamaterial bender.

**Introduction**

In the past few years, the transformation optics [1-3], which provides an intuitive and direct way to control electromagnetic (EM) wave, has received much attention. By associating a desired EM function (e.g., invisibility or bender) with a space of some topological features (e.g., a hole or curvature), one can derive the necessary spatial distribution for the dielectric and magnetic parameters of the related media in the physical space. With the transformation optics, we can design many functional devices, including the reflectionless complex devices such as invisible cloaks [2,4], cylindrical concentrator [5], rotation coating [6], and propagation direction change devices such as beam shifter [7] and beam bender [8]. As one of the interesting devices, the beam bender can change the propagation direction of EM waves; it is examined recently by many authors [8-11]. The EM beam bender proposed by Rahm et al. [8] has anisotropic and spatial-dependent material parameters, which can only be realized with metamaterial technology. By using the equivalent dispersion relations for TE or TM waves, Jiang et al. [9] simplified material parameters for an EM beam bender; however the material is still anisotropic. Within the limit of geometrical optics, Mei and Cui [10] propose an isotropic beam bender. Recently, Landy and Padilla [11] employ the conformal mapping and grid generation techniques to construct an isotropic two-dimensional (2D) bender in the wave optics case; the result is a basically numerical one.

    In this paper, we will propose a new method to design isotropic beam benders. With help of the deformation transformation optics (DTO) [12], the material parameters of an isotropic bender will be derived in an analytical form, in contrast to the numerical optimization solution from conformal mapping technique [11]. The present work also gives another demonstration on how the DTO can be used to simplify the material parameters of a transformation media, in addition to the one given in Ref. [13] to remove the singularity of the transformed material. The paper will be arranged as follows: a design method for isotropic EM beam benders will be presented in section 2; the designed isotropic

bender will be fabricated from stacking dielectric materials and tested in a microwave range. For comparison, the anisotropic beam bender proposed in Ref. [9] will also be fabricated; these will be explained in section 3, and followed by some conclusions.

**The Design Method**

According to the transformation optics, the permittivity and permeability of the designed device have the following expressions [14]

$$\boldsymbol{\varepsilon}' = \mathbf{A}\varepsilon_0\mathbf{A}^\mathrm{T}/\det\mathbf{A}, \boldsymbol{\mu}' = \mathbf{A}\mu_0\mathbf{A}^\mathrm{T}/\det\mathbf{A}, \tag{1}$$

where $\mathbf{A}$ is the Jacobian transformation matrix, $\varepsilon_0$ and $\mu_0$ are the isotropic permittivity and permeability of the original material. Here, we start from an alternative explanation of Eq. (1), which is based on the deformation view of the transformation optics [12]. It is shown that the transformed material parameters can be related to the principle stretches at each point as the following [12].

$$\boldsymbol{\varepsilon}' = \varepsilon_0 \operatorname{diag}[\frac{\lambda_1}{\lambda_2\lambda_3}, \frac{\lambda_2}{\lambda_1\lambda_3}, \frac{\lambda_3}{\lambda_1\lambda_2}], \boldsymbol{\mu}' = \mu_0 \operatorname{diag}[\frac{\lambda_1}{\lambda_2\lambda_3}, \frac{\lambda_2}{\lambda_1\lambda_3}, \frac{\lambda_3}{\lambda_1\lambda_2}], \tag{2}$$

where $\lambda_1, \lambda_2, \lambda_3$ are the three principle stretches of a spatial element during the transformation. To design a 2D bender, consider a rectangular plate with side lengths $ka$ and side width $a$ respectively, as shown in Fig. 1. Under a spatial transformation that makes the plate be transformed into the arc shape (see Fig. 1) with a polar angle $\beta$, it is easily shown that there is only rescaling in $\hat{\boldsymbol{\theta}}$ direction during the transformation, and this stretch can be easily obtained by comparing the arc length of the bender $\beta r$ with its original length $ka$. Therefore, the three principle stretches at each point of the bender have the following forms

$$\lambda_\theta = \beta r/(ka), \tag{3a}$$
$$\lambda_r = \lambda_z = 1. \tag{3b}$$

According to Eq. (2), we get the material parameters of the bender

$$\varepsilon_r = \mu_r = \frac{\lambda_r}{\lambda_\theta \lambda_z} = \frac{ka}{\beta r}, \tag{4a}$$

$$\varepsilon_\theta = \mu_\theta = \frac{\lambda_\theta}{\lambda_z \lambda_r} = \frac{\beta r}{ka}, \tag{4b}$$

$$\varepsilon_z = \mu_z = \frac{\lambda_z}{\lambda_r \lambda_\theta} = \frac{ka}{\beta r}, \tag{4c}$$

These agree with the results given in Ref. [8, 9].

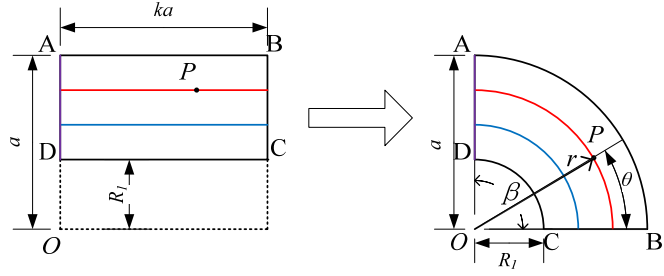

Figure .1. Transformation for a beam bender.

It is important to note that during the bending of the rectangular plate, the deformations along $\hat{\mathbf{r}}$ or $\hat{\mathbf{z}}$ direction will not affect that along $\hat{\boldsymbol{\theta}}$ direction, since these directions are mutually orthogonal to each other. In fact, it can be shown that in this spatial transformation the in-plane stretches $\lambda_\theta$ and $\lambda_r$, as well as the out-of-plane stretch $\lambda_z$ are independent to each other. Therefore, we have many possibilities to simplify the material parameters of the bender by tuning the stretches $\lambda_r$ and $\lambda_\theta$ freely and independently while keeping $\lambda_\theta = \beta r/(ka)$. For example, if we set $\lambda_z = \lambda_\theta = \beta r/(ka)$ and keep $\lambda_r = 1$, the bender can be designed as

$$\varepsilon_r = \mu_r = (\frac{ka}{\beta r})^2, \tag{5a}$$

$$\varepsilon_\theta = \mu_\theta = \varepsilon_z = \mu_z = 1. \tag{5b}$$

Equation (5) coincides with the results of the reducing method by equivalent dispersion relation [9]. However, in this case, the impedance matched conditions in the input and output boundaries are violated [13], there must be some reflections when a beam impinges into the bender. An isotropic bender can be further obtained by setting $\lambda_r = \lambda_z = \lambda_\theta = \beta r/(ka)$, leading to

$$\boldsymbol{\varepsilon} = \boldsymbol{\mu} = \frac{ka}{\beta r}\tilde{\mathbf{I}}. \tag{6}$$

Because this isotropic material is impedance-matched, a perpendicularly incident EM wave will have no reflection in the boundary. Thus, we have constructed a non-reflection beam bender with the isotropic material for full wave, without limit to TE or TM waves. In practice, the gradient of the material parameters can be approximately realized by layering homogeneous isotropic materials. The numerical simulations validate the design based on Eq. (6), as shown in Fig. 2.

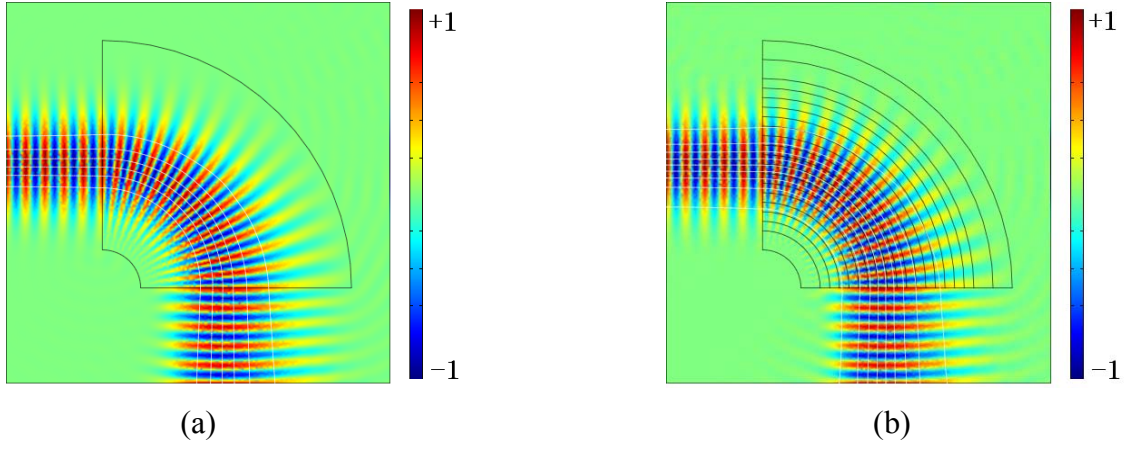

(a)                                (b)

Figure 2. Simulation results for the beam bender with isotropic material parameters Eq. (6) where $a$=1.3m, $k$=1 and $\beta = \pi/2$. The TE waves of frequency 3 GHz incident on left side: (a) Inhomogeneous material. (b) layered-homogeneous material. The white lines indicate directions of the power flow.

However, to make the isotropic permittivity **ε** and permeability **μ** be equal everywhere is very difficult. For practical realization, we set $\lambda_r = \lambda_\theta = \beta r/(ka)$ and keep $\lambda_z = 1$, then we obtain

$$\varepsilon_r = \mu_r = 1, \varepsilon_\theta = \mu_\theta = 1. \tag{7a}$$

$$\varepsilon_z = \mu_z = (\frac{ka}{\beta r})^2. \tag{7b}$$

Eq. (7) suggests the following dielectric bender for a TE wave

$$\varepsilon_z = (\frac{ka}{\beta r})^2, \tag{8a}$$

$$\mu_r = \mu_\theta = 1. \tag{8b}$$

Here the magnetic permeability $\mu$ is isotropic value. Although small reflections exist for this bender as that of Eq. (5), it has strong practical advantages for using isotropic traditional materials.

For the TM incident wave, the electric permittivity $\varepsilon$ is isotropic value and the Eq. (7) can be written as

$$\mu_z = (\frac{ka}{\beta r})^2, \tag{9a}$$

$$\varepsilon_r = \varepsilon_\theta = 1. \tag{9b}$$

**Experiments**

**Metamaterial beam bender.** The beam bender with the reduced parameters proposed by Ref. [9] is firstly fabricated for comparison. The metamaterial layer based on split-ring resonators (SRR) is designed and fabricated with the printed circuit technology. In the sample preparation, the substrate of sample we adopt is the printed circuit board (Rogers® RT/duroid 5880) with the relative permittivity $\varepsilon$ =2.20 and the thickness of 0.25mm. Because of constraints of the layout, we choose the seemingly arbitrary radial values a = 138 mm and b = 93.4 mm. Then we also choose a rectangular unit cell with dimensions $a_\theta = a_z = 10/3$ mm and $a_r = 10/\pi$ mm. The others geometric parameters, the radius of the corners $r = 0.1$ mm, the width of SRR metal wire $w = 0.2$ mm and the distance between two splits $x = 0.1$ mm. The resulting dimension properties are plotted in Fig.3 (a). The desired permeability of each layer is realized with SRR unit cells by tuning one of its geometrical parameters: the length of the split (s). Using commercial, full-wave, finite-element simulation software (Ansoft HFSS), we perform a series of scattering (S) parameter simulations for the SRR unit cells over a discrete set of the parameter s covering the range of interest. A standard retrieval procedure is then performed to obtain the effective material properties $\mu_r$ from the S parameters. The permeability profile is realized by 15 layers designed at 8.5GHz, as shown in Fig.3 (b), and the sample of the metamaterial bender we fabricated is shown in Fig.4.

To examine the functionality of the beam bender, we have built up an electric-field mapping system shown in Fig. 5, which is similar to the setup in Ref. [4]. The electric fields measured on the surface of the metamaterial layers are shown in Fig. 6 (a). It is seen that the designed beam bender indeed bends the wave propagation direction in the desired way at the working frequency of 8.5 GHz. However, the bender losses its function when the frequency is changed to 10.5 GHz, as shown in Fig.6 (b). This is expected since the metamaterials with the SRR structures are restricted to a narrowband and large-loss normally. In recent years, the bandwidth of the novel metamaterials is theoretical and experimental proved to be a broad frequency band by some special structure [15].

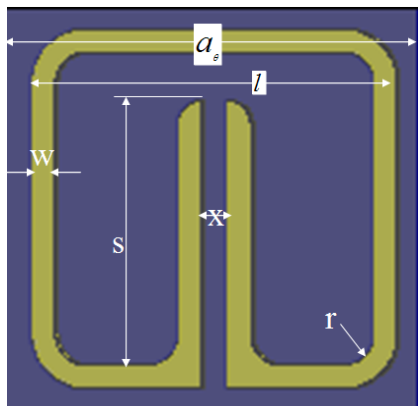    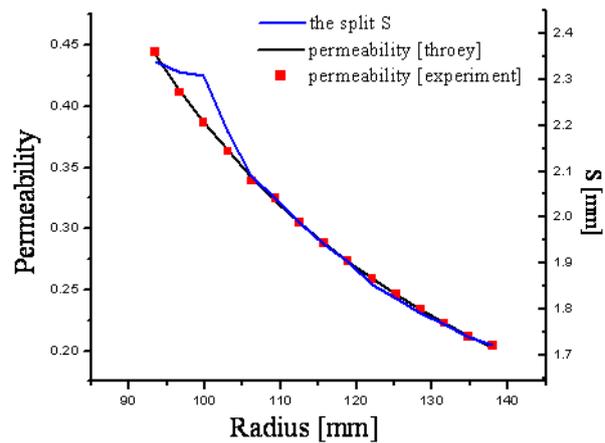

(a)                                          (b)

Figure 3. (a) The SRR model for the metamaterial. (b) The parameters of the metamaterials bender.

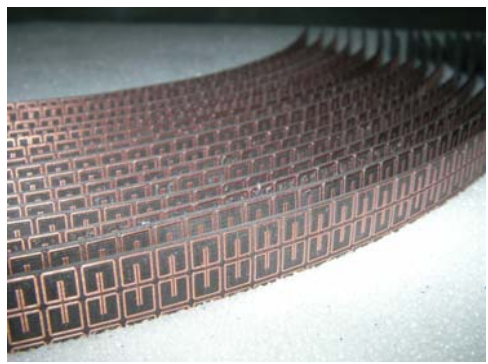

Figure 4. The sample of the metamaterial bender.

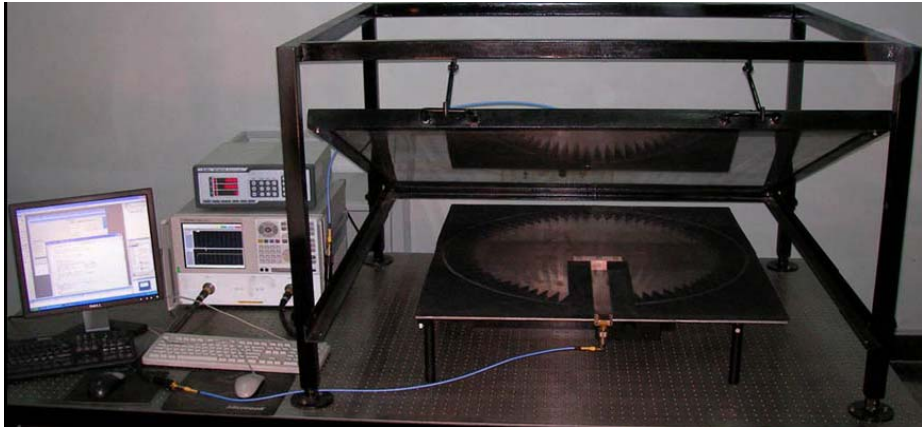

Figure 5. The electric-field scanning system.

**Isotropic dielectric beam bender**. The isotropic bender suggested by Eq. (8) has only one design parameter that depends on the radial coordinate; it can be fabricated by stacking dielectric layers with different permittivity. We choose the value of the periphery radius $a = 100$ mm, $\beta = \pi/2$, $k^2 = 5$ in Eq. (8). The high-frequency printed circuit boards (PCB, Taconic® and Rogers®) are used to fabricate the desired isotropic beam bender by adjusting the correlation with the layer thickness and the desired permittivity $\varepsilon_z$ which is from 2.08 to 3.50. Consideration with the influence of the metal on the PCB, we corrode the metal on the PCB's surface and cut the substrate into many thin slices with the height of 10mm as our basic sample materials. The sample and the parameters of each layer are shown in Fig. 7. The numerical simulation is also employed to validate the design model, and the result is shown in Fig. 8 with a frequency 8.5 GHz. Since the bender is made of traditional materials, so it is expected to have a broadband and low-loss. In whole system, the main loss comes from the bender materials and it is much loss for PCBs, which the dissipation factor (tan δ) is about 0.002 at 10 GHz and 23 °C. By the way, the thickness of the bender sample is 32.5 mm along the radial direction, and the width of wave-guide (model: WR-90) which is connected to the Vector Network Analyzer (Agilent® E8362B) is 22.86 mm. The frequencies range of the WR-90 wave-guide is from 8.2 GHz to 12.4 GHz, so microwaves are introduced through this X-band coax-to-waveguide adapter that is attached to the lower plate in the electric-field mapping system and are incident on the sample, which rested on the lower plate and was nearly of the same height (10 mm) as the plate separation

With the help of the electric-field mapping system, the electric field distributions of the bender are measured at the different frequencies 8.5 GHz, 10.5 GHz and 12 GHz, respectively, they are shown separately in Figs. 9 (a), (b) and (c). We can see that the designed layered laminates can bend the EM wave almost perfectly from 8.5GHz to 12GHz, exhibiting clearly the wider bandwidth compared to the metamaterial beam bender. In addition, because of adopting the traditional materials, we can also see there is not loss or tiny loss. It can be demonstrated through the color-bar of the electric field figures. The electric field at 12GHz is not perfect; it is probably due to the variation of the electromagnetic parameters for the board material with the frequency.

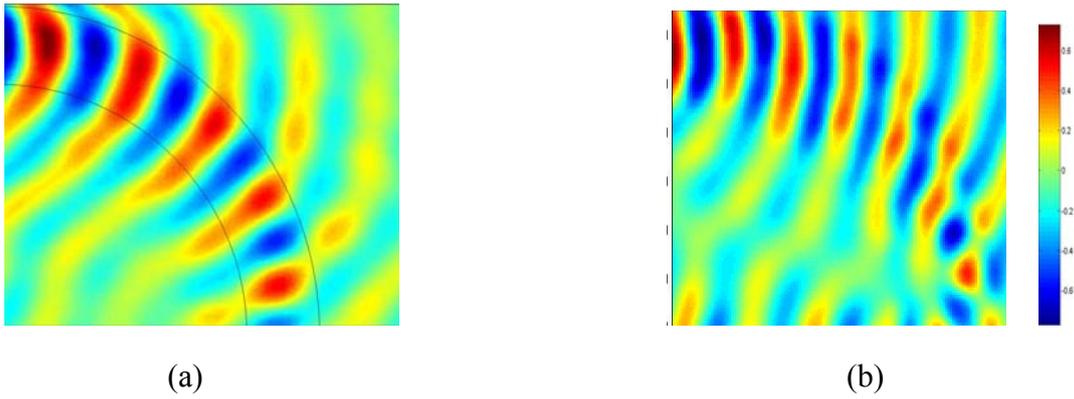

Figure 6. The experimental results for electric field distribution (z component) with the metamaterial bender at (a) 8.5GHz and (b) 10.5GHz, respectively

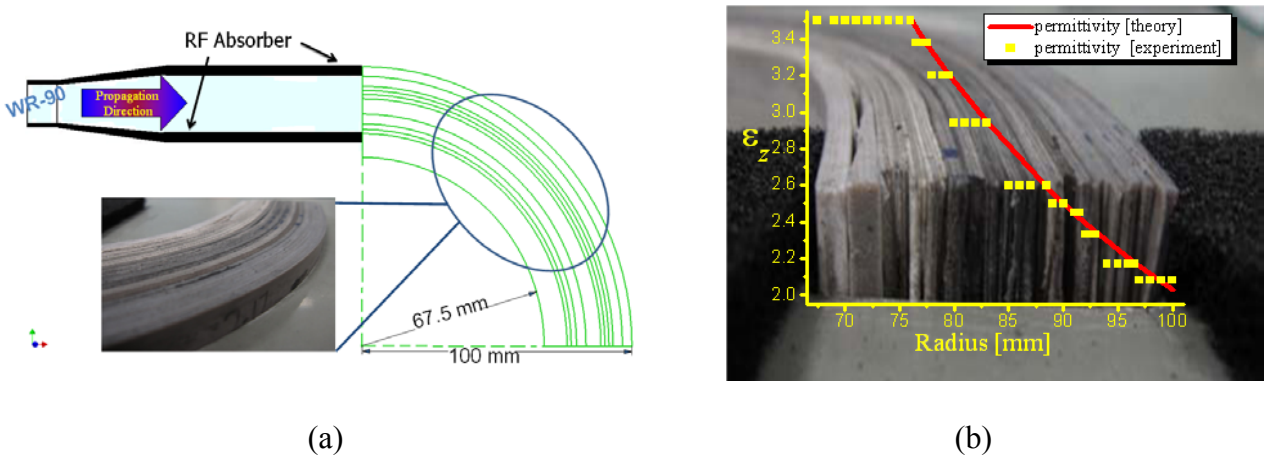

Figure 7. (a) The experimental scheme and the bender sample made of stacked dielectric layers. (b) The parameters of the dielectric bender

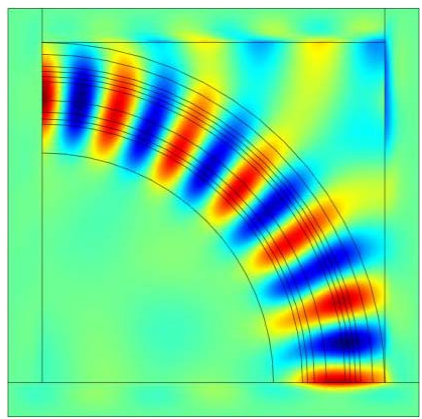

Figure 8. Simulation results for electric field distribution (z component) with the layered dielectric materials.

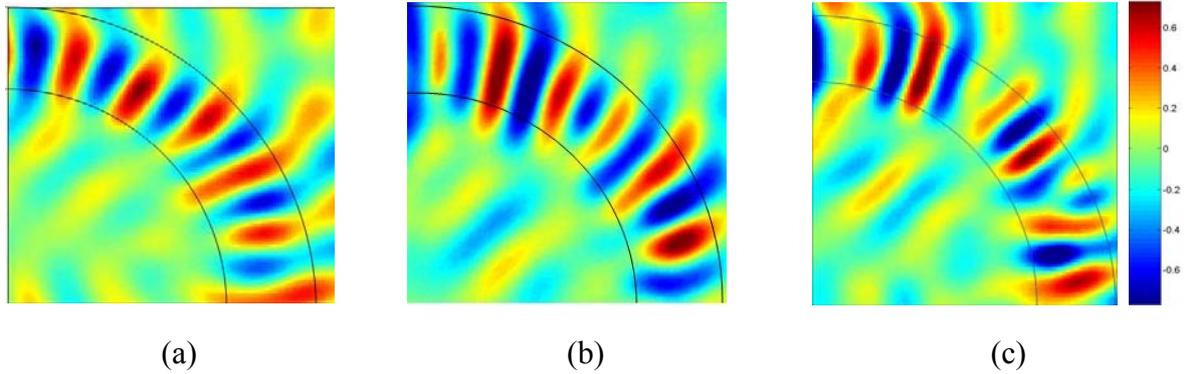

Figure 9. The experimental results for electric field distribution (z component) with isotropic bender at the frequencies (a) 8.5 GHz, (b) 10.5GHz and (c) 12GHz, respectively.

**Conclusions**

Based on DTO, we have presented a method to design and simplify EM beam benders. According to continuum mechanics [16], any deformation of a region, described by the Jacobian transformation matrix or deformation gradient tensor **A**, will have principal stretches mutually orthogonal to each other on every point. In general, the principal stretches can be obtained by the square roots of eigenvalues of the left Cauchy-Green deformation tensor $\mathbf{B} = \mathbf{A}\mathbf{A}^T$, thus the proposed idea of adjusting principal stretches is in fact adaptable in simplifying arbitrary shaped devices, as shown in Ref. [13]. Another numerical optimization method for design simple EM beam benders also based on deformation theory can be found in Ref. [17]. However, for the regular arc-shape bender, the principal stretches are obvious and we can simplify the bender design in analytical ways, which are more useful in practice than the numerical ways. By comparing the change of a line segment during the transformation, necessary material parameters for the bender can be easily obtained. Since the three principal stretches during the spatial transformation of a beam bender are independent, by adjusting the principal stretches, the benders proposed in the literatures can be recovered as a special case. We have also designed an isotropic bender, which can be realized by layering isotropic tradition materials, the proposed bender is broadband and has low loss compared to the metamaterial bender. A metamaterial bender based on SRRs and an isotropic bender with layered dielectric materials are both fabricated and measured, the results confirm the theoretical design. For the isotropic bender with the traditional parameters of the material, it is much easy to design and fabricate in the wide material choice in comparison with the metamaterials.

**Acknowledgements**

This work is supported by the National Natural Science Foundation of China (Grant Nos. 90605001, 10702006, and 10832002), and the National Basic Research Program of China (Grant No. 2006CB601204).